# Diagnosis of electron density and temperature by using collisional radiative model in capacitively coupled Ar plasmas Ⅱ: two-dimensional distributions

**Jidun Wu[1], Hao Zheng[1], Yanfei Wang[1], Fengzhu Zhou[1] and Xiaojiang Huang[1,2,3] iD**

[1] College of Science, Donghua University, Shanghai 201620, China
[2] Member of Magnetic Confinement Fusion Research Centre, Ministry of Education of China, Shanghai 201620, China
[3] Textiles Key Laboratory for Advanced Plasma Technology and Application, Shanghai 201620, China

E-mail: huangxj@dhu.edu.cn

## Abstract

Effects of radio-frequency power and driven frequency on the two-dimensional (axial and radial) distributions of electron density and temperature were experimentally investigated in low pressure capacitively coupled argon plasmas. The intensity profiles of 696.5 nm and 750.4 nm emission lines were detected by employing a spatially resolved diagnostic system, which consists of a charge coupled device (CCD) and bandpass interference filters. The two-dimensional distributions of electron density and electron temperature were calculated from the spatial distributions of emission intensities via a collisional radiative model (CRM). It is found that the axial and radial distributions of electron density are more uniform at a lower RF power. The axial uniformity of electron density is better at a lower driven frequency, while the radial profiles of electron temperature is flatter at a higher excitation frequency. In all the cases, the electron temperature is extremely uniform in the bulk plasma. Moreover, a mode transition from the α to the γ mode is observed with the increase of input RF power at 13.56 MHz, which causes a significant increase of electron density and an abrupt decrease of electron temperature.

## 1. Introduction

Radio-frequency (RF) capacitively coupled plasmas (CCPs) have been extensively utilized for materials surface processing in microelectronics industry over the past few decades [1, 2]. Due to the lack of reliable experimental evidence for electron behaviors, the discharge phenomena in low pressure CCP discharges are still not completely understood, such as discharge structure, electron power absorption modes and mode transitions between them. Therefore, simple and powerful diagnostic methods are necessary to understand basic and meaningful principles behind these experimental phenomena.

In order to better comprehend the underlying physics of CCPs, some numerical simulations and experimental researches have been made, from which the spatial distributions of plasma parameters have been observed. Zhang *et al* [3] demonstrated that the electrical asymmetry effect can control the spatial distributions of plasma density using a two-dimensional particle-in-cell/Monte Carlo (PIC/MCC) simulation. Liu *et al* [4] studied the plasma non-uniformity in electron heating in a CCP reactor with a dielectric side wall by both a two-dimensional PIC/MCC program and an analytical model. They clarified that the enhanced Ohmic heating with decreasing electron density towards the electrode edge results in the non-uniformity in electron heating. Later, they investigated the effect of the excitation frequency on the time-averaged electron density adopting the same PIC/MCC model [5]. They found that the radial plasma uniformity can be improved by increasing the driven frequency.

However, the previous experimental studies were mainly based on one-dimensional diagnostic systems, such as Langmuir probe, floating double probe and hairpin probe. Zhao *et al* [6] employed a Langmuir probe to diagnose the axial profiles of electron density and electron temperature. They observed that electron density near the powered electrode is higher than that near the grounded electrode in pure argon discharges. Based on a synergetic use of floating double probe and two-dimensional self-consistent electrostatic fluid model, Liang *et al* [7] presented that the radial uniformity of plasmas is better at low RF power and large electrode gap. By using a hairpin probe, Liu *et al* [8] studied the axial distribution of electron density in neon plasmas driven at single frequency. They discovered that the electron density is more uniform in the bulk plasma when RF voltage is low. In a later paper [9], Liu *et al* showed that the peak value of plasma density is very close to the powered electrode both in single-frequency and dual-frequency argon discharges, as a result of the ionization maximum located at the plasma-sheath boundary in front of the powered electrode.

In recent years, some researchers [10-14] developed the phase resolved optical emission spectroscopy (PROES) to diagnose the dynamics of highly energetic electrons. By measuring the emission intensity from a specifically chosen neon state (Ne-$2p_1$) at 585.5 nm, the spatio-temporal electron-impact excitation rate from the ground state into the Ne-$2p_1$ state can be obtained through a time dependent model. Schulze *et al* [10, 11] systematically reviewed the method of PROES and revealed a novel electron power absorption mode in electronegative plasma discharges, which was described as the Drift-Ambipolar mode. Liu *et al* [12-14] discovered another electron power absorption mechanism, called the striated mode in the electronegative gas ($CF_4$), using the PROES and PIC/MCC simulations.

PIC/MCC simulations, electrical probes and PROES are widely used for diagnosing the spatial distribution of CCPs, but few researchers have focused on the two-dimensional spatial distribution of plasma parameters experimentally, e.g., electron density and electron temperature. Two-dimensional PIC/MCC simulations usually take a long time to calculate and obtain spatial profiles of various plasma parameters. In addition, due to the limitation of Langmuir probe or hairpin probe, it is hard to achieve two-dimensional spatial resolution. Zhu *et al* [15] proposed a spatially resolved optical emission spectroscopy (OES) diagnostic system free of chromatic aberrations for micro-plasma. The spatial distribution of time-averaged effective electron temperature was estimated with a CRM. Sanghoo *et al* [16] diagnosed the two-dimensional distribution of electron temperature based on neutral bremsstrahlung emissivities in atmospheric pressure plasmas, by utilizing a digital camera and optical interference filters. However, neither of them gave the spatial distributions of electron density and electron temperature simultaneously. As such, the purpose of this paper is to introduce a two-dimensional experimental diagnostic method for CCPs and discuss two cases of its applications.

In this work, we present a spatially resolved diagnosis system for low pressure argon plasma parameters measurement by combining a charge coupled device (CCD) camera with two optical bandpass interference filters. The diagnostic system is operated on a capacitively coupled argon plasma at different driven frequencies and RF power. The two-dimensional spatial distributions of time-averaged effective electron temperature and electron density are simultaneously obtained from the intensity distributions of 696.5 nm and 750.4 nm emission lines through a CRM, and we discuss the effects of input power and driven frequency on their spatial distributions. Besides, the mode transition from the α to the γ mode are also illustrated, by analyzing the dependence of electron density and electron temperature on the RF power. Note that the power mentioned in this work is just the applied RF power, not a measured RF power absorption by plasma practically.

The paper is structured as follows: In Section 2, the experimental setup and diagnostic systems are described. Then the calibration method between spectrometer and CCD camera, which is used to measure the spatially resolved optical emissions, is introduced briefly. In Section 3, experimental results are presented and discussed. Finally, conclusions are drawn in Section 4.

## 2. Experimental setup and diagnostic technique

*2.1 Experimental setup*

Figure 1 shows the schematic diagram of the CCP reactor and the diagnostic system. Two electrodes are 5 cm in diameter and placed vertically in the reactor. Note that the left electrode can move freely when reactor is not under a vacuum condition, thus the adjustable range of electrode gap is around 0-50 cm. But in this work, the electrode gap is fixed at 3 cm. RF power is applied to the right electrode via an impedance matching box, and left electrode is grounded. Both electrodes are made of aluminum, while the chamber wall is made of quartz. When the gate valve is open, the reactor can be evacuated through a rotary pump with a pumping speed of 1400 r/min, and the base pressure of the chamber is $10^{-1}$ Pa. OES is monitored by an Avaspec-2048TEC in the absolute irradiance mode, with its spectral resolution of 0.13 nm, and the range of wavelength from 200 nm to 900 nm. Instead of being a major diagnostic tool, the spectrometer is utilized to calibrate the emission intensities detected by a CCD camera in this paper.

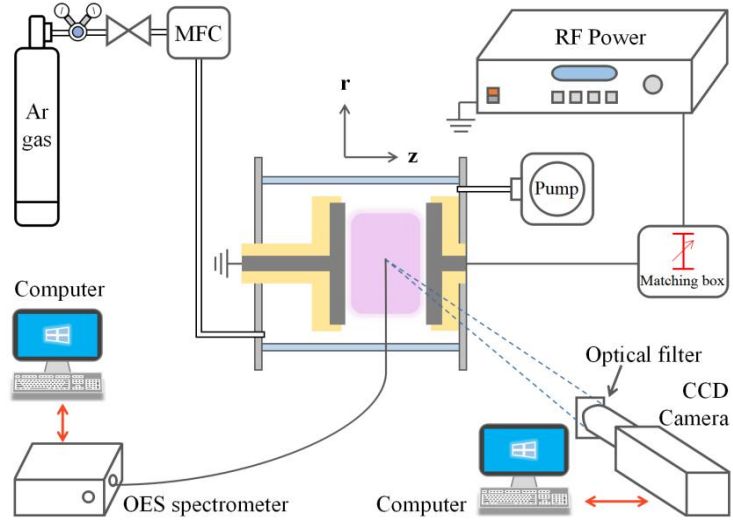

**Figure 1.** Schematic of the CCP reactor and diagnostic system.

*2.2 Diagnostic technique*

The sensor size of the CCD camera (Basler acA 1300-60 gm) is 6.8 × 5.4 mm and the resolution is 1282 × 1026 px. Therefore, the pixel size of the CCD is 5.3 × 5.3 μm. The CCD camera using a lens with a fixed focal length of 50 cm, focuses on the center of discharges and the distance between them is fixed at 75 cm. The emission intensities of 696.5 nm and 750.4 nm are recorded by inserting optical bandpass interference filters with different center wavelengths in front of the CCD camera. The central wavelength of two filters are 700 and 750 nm respectively, and the full width at half maximum of the filters is 10 nm. Therefore, two emission lines in capacitively coupled argon plasma discharges can be detected with these filters, respectively. Although the intensity of 751.5 nm emission line is also detected due to the filter with 10 nm half width, the emission intensity ratio of 750.4 nm and 751.5 nm is almost constant under the conditions of this article, which is examined by OES measurement. Hence, the 751.5 nm emission has a limited influence on the calibration coefficient of the emission intensity at 750.4 nm.

When discharges were operated, CCD camera and spectrometer were simultaneously used to measure the emission intensities of 696.5 and 750.4 nm under same discharge conditions. We chose 40 × 40 = 1600 pixels in an image of CCD and averaged their luminance values as a data point, which is approximately corresponding to the 3 × 3 mm space collected by the spectrometer. Therefore, the two calibration coefficients of the above emission intensities detected by spectrometer and CCD camera could be determined respectively. It is confirmed that the calibration coefficients almost linearly increase as the increase of exposure time.

A total of 16 × 10 = 160 data points of the emission intensities of 696.5nm and 750.4 nm, respectively, are selected to calculate the spatial profiles of electron density and electron temperature with a CRM. More details about the CRM and the calculation method can be found in [our paper I]. A simple example is presented in figures 2. In this case, the working pressure $p$ is 20 Pa，the driven frequency $f$ is 13.56 MHz and the input power $P_{rf}$ is set to 140 W. The radial direction is the $r$-axis and the axial direction is the $z$-axis, the grounded electrode is located at $z = 0$ cm, while the powered electrode is at $z = 3$ cm. This applies to all the figures in this work. Figure 2(a) and (b) show the filtered emission profiles of 696.5 and 750.4 nm. The exposure times of figure 2(a) and (b) are 7500 and 1000 μs respectively. It can be seen that the radial emission distributions are symmetric, while emission intensities peak near the plasma sheath boundaries in the axial direction. The spatial distributions of electron density and electron temperature calculated with a CRM are presented in figure 2(c) and (d). Obviously, both radial profiles of electron density and electron temperature are very symmetric. In the axial direction, the maximum value of electron density is localized at the position closed to powered electrode. The electron temperature exhibits maxima in front of two electrodes and a good uniformity within the bulk plasma.

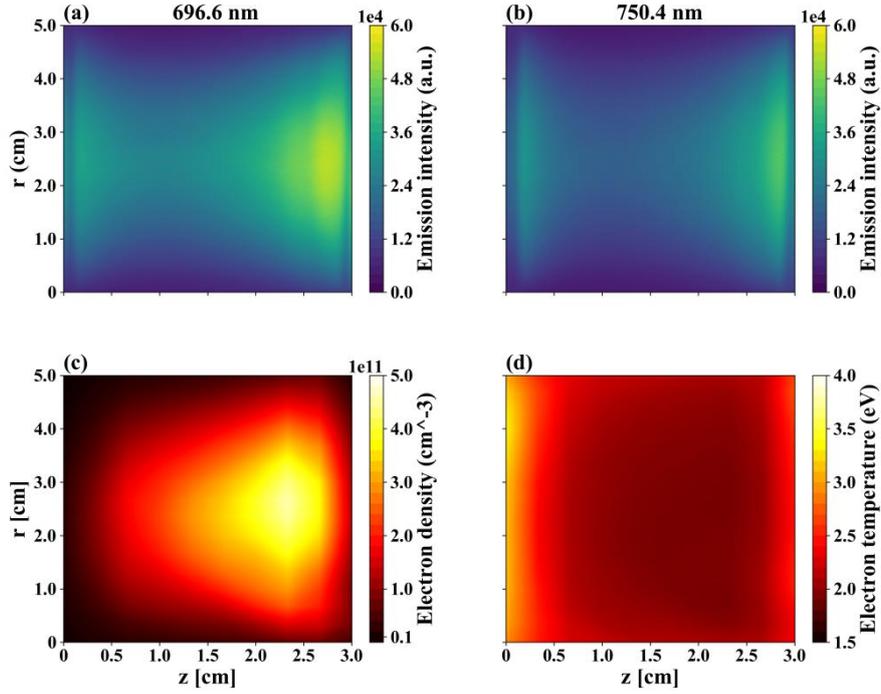

**Figure 2.** The filtered emission distributions of (a) 696.5 nm, (b) 750.4 nm. (c) electron density and (d) electron temperature obtained with a CRM. Note that $z = 0$ indicates the grounded electrode and $z = 3.0$ cm indicates the powered electrode. This applies to all the figures in this work. Discharge conditions: $p = 20$ Pa, $f = 13.56$ MHz, $P_{rf} = 140$ W.

Compared with spectrometers, there are some advantages applying a combination of a CCD camera and two optical interference filters to diagnose plasma discharges. It provides easy access to one-dimensional distributions of electron density and electron temperature, e.g., axial and radial profiles. More importantly, it can achieve two-dimension spatially resolved diagnostic of plasma parameters and even temporally resolved by substituting an intensified CCD (ICCD) for the CCD. Besides it is assumed that plasmas are optically thin, thus an Abel inversion is not considered for processing the spatial distributions of emission intensities.

## 3. Results and discussion

### 3.1 Two-dimensional distributions of electron density and temperature with changing power

In this section, we demonstrate the influence of input power on the spatial profiles of electron density and electron temperature. The mode transition from the α to the γ mode also is discussed. The RF power ranges from 40 W to 140 W and the excitation frequency is 13.56 MHz. The argon gas pressure, is set to 20 Pa and the electrode gap is fixed at 3 cm.

Figures 3(a)-(c), (d)-(f) and (g)-(i), show the spatial distributions of emission intensity at 750.4 nm, electron density and electron temperature, respectively. Due to the similar trend of emission intensity of 696.5 nm and 750.4 nm with RF power, we just analyze the spatial distribution of emission intensity at 750.4 nm. It can be seen from figures 3(a)-(c) that the emission intensity at 750.4 nm, rises with the increase of input power in the whole discharge region. And emission intensity becomes stronger near the sheaths in front of two electrodes, it indicates that the excitation is effectively enhanced at these regions when input power is relatively high.

It should be noted that the electron density varies greatly with power in figures 3(d)-(f), thereby the range of colorbar is not uniform at different power and each graph takes the range of 26 times the maximum value of the minimum value. When input power gradually increases from 40 W to 140 W, the electron density significantly rises in the whole discharge regions, especially in the position that is close to powered electrode. Both axial and radial uniformities of electron density at a relative low power are better than that at a higher power. The diameter of two electrodes used in this work is small (5 cm), thus the radial plasma uniformity is not the main point of this article. The dependence of electron temperature on input power are displayed in figures 3(g)-(i). In contrast to electron density, the electron temperature declines clearly with increasing input power. The spatial profile of electron temperature is very uniform in the bulk plasma in three cases.

In order to further analyze the axial plasma uniformity, the evolutions of axial distributions (at $r = 2.50$ cm) of electron density and electron temperature with the applied RF power, are shown in figure 4. In Figure 4(a), it can be found that the electron density monotonically increases with increasing RF power. The axial profile of electron density is in parabolic shape and the maximum is close to the powered electrode. The similar axial distribution of electron density was also observed in the experiments [6, 9] and simulations [17, 18]. Due to the asymmetric discharge between two electrodes, the powered electrode exists a dc self-bias [8]. Compared with the grounded electrode, thus, the electron density is higher near the driven electrode, where the ionization maximum occurs. The maximum of electron density is twice the minimum of electron density in the axial direction at 40 W, while the ratio of maximum and minimum reaches 13 at 140 W. The electron density is more uniform at the low RF power than that at the high RF power consequently. The central electron density increases almost 54 times between 40 and 140 W. This is attributed to markedly enhanced ionization caused by secondary electrons, which contributes to more rapid increase of electron density in front of powered electrode.

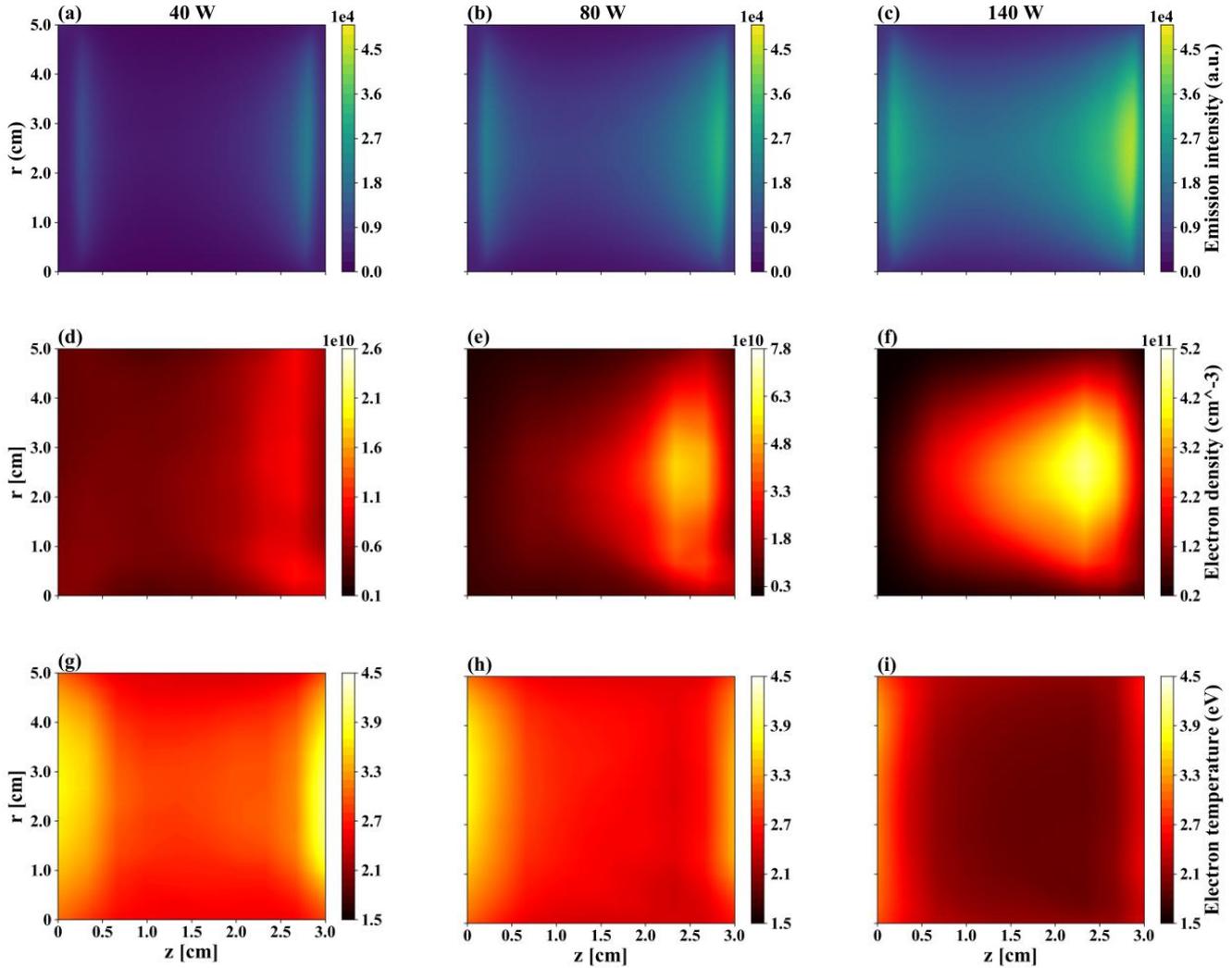

**Figure 3.** The spatial profiles of (a)-(c) emission intensity, (d)-(f) electron density and (g)-(i) electron temperature at various input power. Discharge conditions: $p$ = 20 Pa, $f$ = 13.56 MHz, and the RF power ranges from 40 W to 140 W. Note that here the intensity of emission line chosen is at 750.4 nm.

In figure 4(b), the electron temperature decreases with the increase of RF power. The axial distribution of electron temperature exhibits a saddle type under all the power conditions. The saddle shape of electron temperature is similar to the results of simulations [17, 18]. Generally, electron temperature is uniform within the bulk plasma. Because the electron has a very tiny mass, it can response to instantaneous electric field and gain substantial energy from electric field directly. The energy gained from the electric field in the bulk plasma is used to compensate for the energy loss of the electron collisions, thus the electron temperature near two electrodes is higher [17]. In addition, at a lower power, the electron temperature is higher near the driven electrode than that near the grounded electrode. However, the tendency gradually changes with increasing RF power. The electron temperature near the powered electrode rapidly drops and is lower than the electron temperature near the opposite electrode at a relatively high input power. The reason is that the electron density near the powered electrode is so high that the collisions between electrons and neutral particles become extremely frequent, which leads to an increase of electron energy loss.

The dependence of electron density and electron temperature at the position ($r$ = 2.50 cm and $z$ = 1.65 cm) near the discharge center with the input power is shown in figure 5. Both electron density and electron temperature occur significant changes with increasing RF power. In the range of 40-80 W, the electron density slightly increases as the power varied and then rapidly increases after RF power exceeding 80 W. This indicates the occurrence of mode transition from the α to the γ mode. The sheath voltage increases with the increase of RF power. The secondary electrons are emitted from the surface of powered electrode due to ion and neutral particles bombardment. These secondary electrons gain substantial energy from sheath electric field when traversing the sheath region, contributing to significant avalanche ionization at the plasma-sheath boundary. The electrons generated by avalanche ionization enter the bulk region and collide with the background gas frequently, which causes a sharp increase of electron density. The electron power absorption mode switches from the α mode dominated by sheath expansion to the γ mode dominated by secondary electron emission.

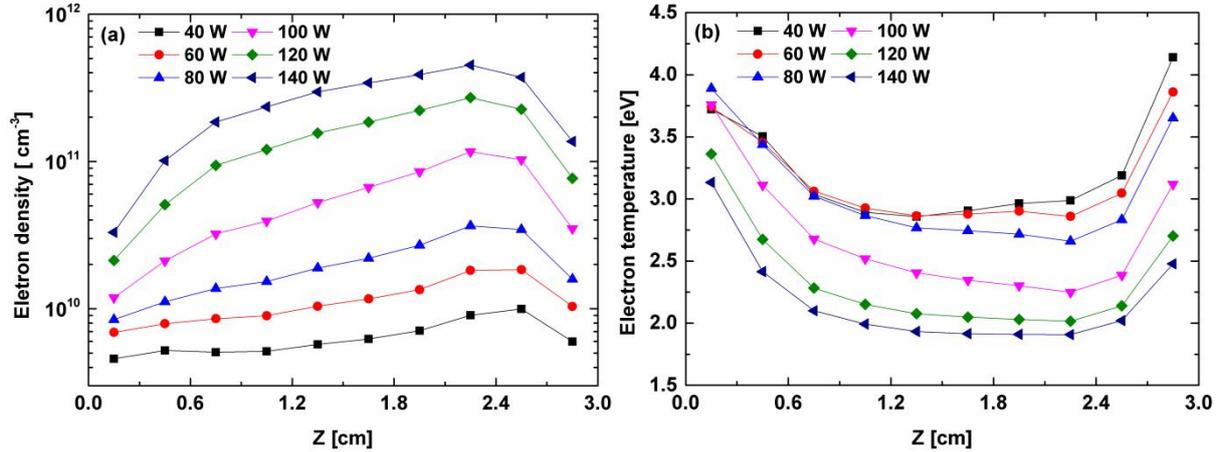

**Figure 4.** The axial distributions of (a) electron density and (b) electron temperature at different RF power. Discharge conditions: $p$ = 20 Pa, $f$ = 13.56 MHz, and the RF power is varied between 40 W to 140 W.

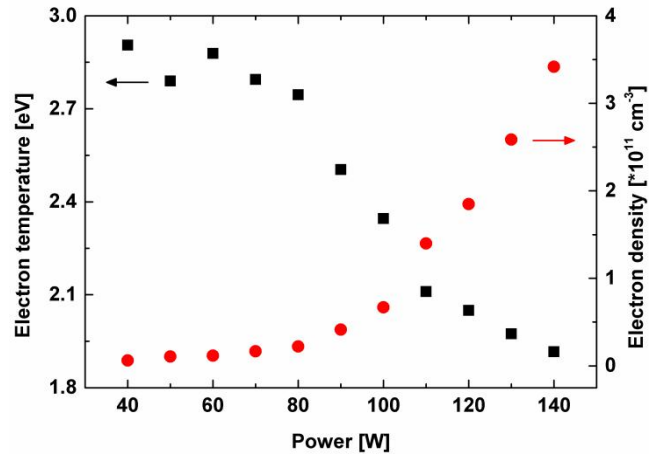

**Figure 5.** The electron density and the electron temperature near the discharge center ($r$ = 2.5 cm, $z$ = 1.65 cm) as a function of the RF power in argon discharges. Discharge conditions: $p$ = 20 Pa, $f$ = 13.56 MHz, and the input power is varied between 40 W and 140 W. The distinct changes in electron density and electron temperature with increasing power indicate that the mode transition from the α to the γ mode.

The transition point also can be defined by the steepest $T_e(P_{rf})$ dependence [19], corresponding to a transition $P_{rf}$ of 80 W. It can be seen that the electron temperature slightly decreases with increasing power at first and abruptly decreases when RF power exceeds the transition point. The reason is the dramatically increased collision frequency between electrons and neutral atoms. Same trends of electron density and electron temperature with current density or input power have been presented elsewhere [19-21]. In our work, the order of magnitude of electron density is almost consistent with the results of Godyak *et al* [19, 20] in argon and helium capacitively discharges, but higher than that in $O_2$ plasma operated at 100 mTorr [21]. It is due to the presence of negative ions, which causes a reduction of electron density to maintain the quasi-neutrality in the bulk plasma in the negative plasma discharges.

*3.2 Two-dimensional distributions of electron density and temperature under changing frequencies*

In this section, the effect of driven frequency on the spatial distributions of electron density and electron temperature is illustrated. The excitation frequency is 2 MHz, 13.56 MHz and 27.12 MHz respectively. The electrode gap is 3 cm and the input power is kept constant, 80 W. The argon gas pressure is fixed at 10 Pa, in order to ensure that discharges are operated in the same electron power absorption mode (α mode) when analyzing the spatial distributions of plasma parameters.

Figures 6(a)-(c), (d)-(f) and (g)-(i), show the spatial distributions of emission intensity at 750.4 nm, electron density and electron temperature, respectively. Because of the similar evolution of emission intensity of 696.5 nm and 750.4 nm with frequency, we just discuss the spatial distribution of emission intensity at 750.4 nm. One can see from figures 6(a)-(c) that the intensity of emission line at 750.4 nm, rises with increasing driven frequency and the spatial distribution is more uniform at low frequency. Moreover, the light intensity near the sheaths visibly enhances when discharge is operated at 27.12 MHz.

Similarly, the electron density changes dramatically with frequency increases in figures 6(d)-(f), hence the range of colorbar is not uniform at different frequencies and each graph takes the range of 10 times the maximum value of the minimum value. In figures 6(e)-(f), when driven frequency increases from 2 MHz to 27.12 MHz, the electron density evidently increases in the whole discharge space. In capacitively coupled discharges operated with fixed power, the fraction of RF power depositing into the electrons rises with the increase of driven frequency, leading to a sharp enhancement in plasma density [1]. It is obvious that the electron density gradient is lager at 27.12 MHz, hence the axial plasma uniformity is better at 2 MHz with a smaller gradient of electron density. By comparing the electron temperature at different frequencies in figures 6(g)-(i), we can see that electron temperature decreases with increasing driven frequency. As the frequency gets lower, the sheaths get thicker, and the electric field can effectively penetrate into deeper of the bulk plasma. In three cases, the electron temperature in the bulk region shows an excellent uniformity. Nevertheless, the radial distribution of electron temperature is more uniform at 27.12 MHz due to a small gradient of electron temperature.

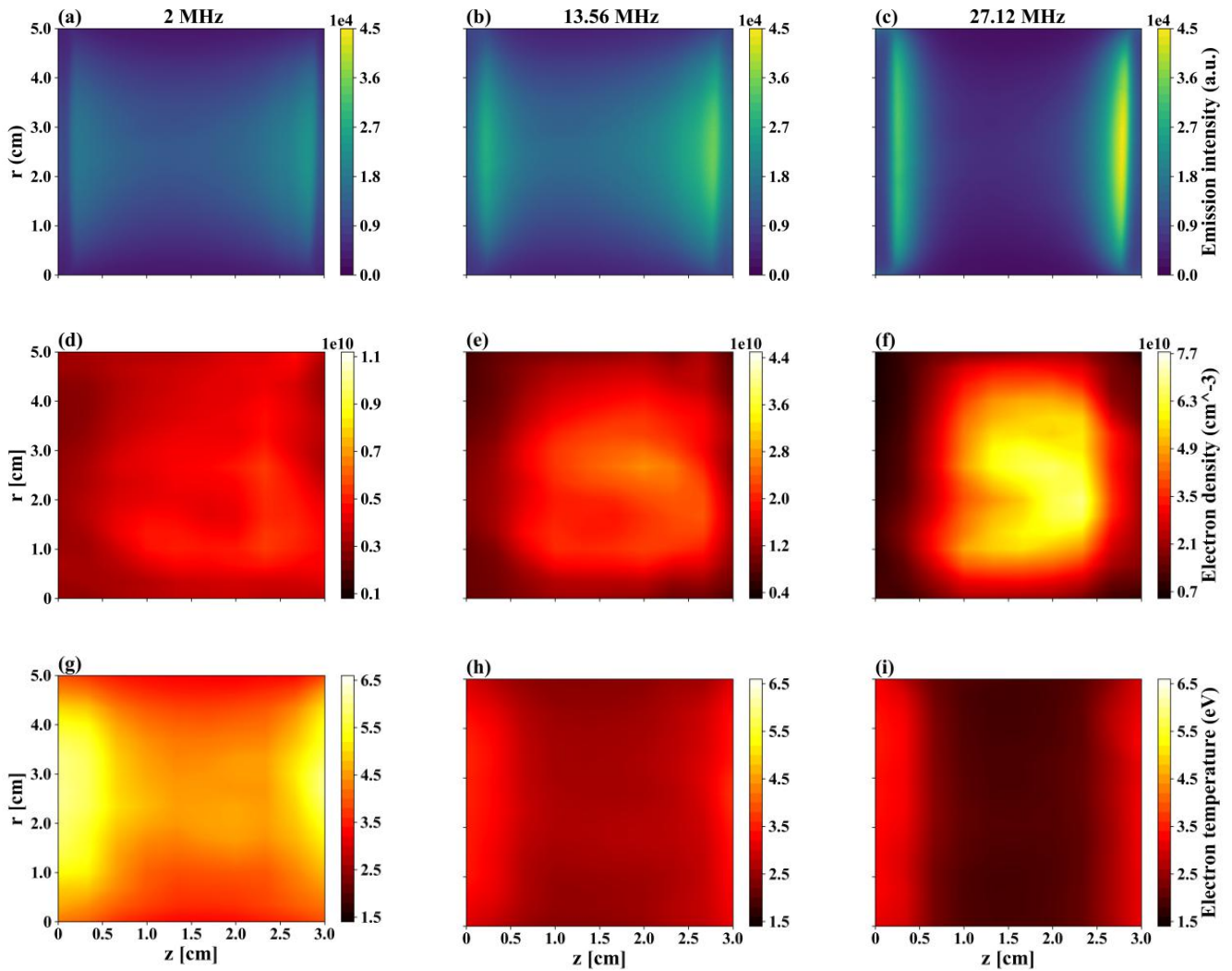

**Figure 6.** The spatial profiles of (a)-(c) emission intensity, (d)-(f) electron density and (g)-(i) electron temperature at different driven frequencies. Discharge conditions: $p$ = 10 Pa, $P_{rf}$ = 80 W. The excitation frequency is 2 MHz, 13.56MHz and 27.12 MHz respectively. Note that here the intensity of emission line chosen is at 750.4 nm.

The axial distributions of electron density and electron temperature as a function of the RF frequency (at $r = 2.50$ cm) are shown in figures 7 under the same conditions as in figures 6. In figure 7(a), it can be found that the electron density increases when increasing frequency. The axial profile of electron density shows a parabolic shape at all frequencies, and the maximum is slightly close to the powered electrode because of asymmetric discharge structure. The ionization maximum occurs when near the powered electrode, instead of grounded electrode. The ratio of maximum and minimum of electron density in the axial direction is 1.83 at 2 MHz, while the electron density increases almost 5 times between minimum and maximum at 27.12 MHz. For 27.12 MHz with a large gradient of electron density in the axial direction, it indicates that the axial uniformity is worse at a higher frequency. In the bulk plasma, the electron density is more uniform with a small gradient at a lower frequency. In figure 7(b), the electron temperature drops with the increase of driven frequency. The axial distribution of electron temperature exhibits a saddle type under these frequency conditions, which is similar to figure 4(b).

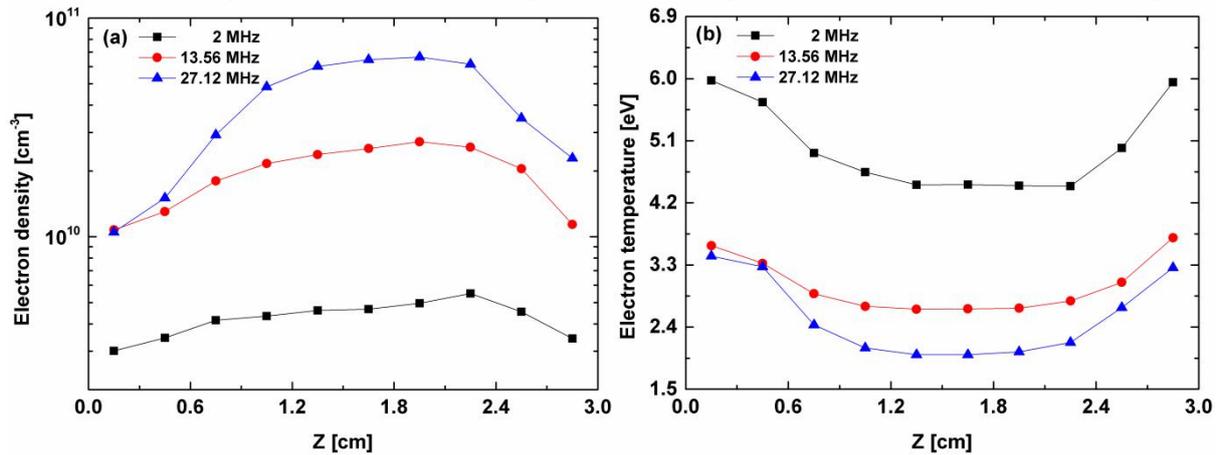

**Figure 7.** The axial distributions of (a) electron density and (b) electron temperature at different RF frequencies. Discharge conditions: $p = 10$ Pa, $P_{rf} = 80$ W. The driven frequency is 2 MHz, 13.56 MHz and 27.12 MHz respectively.

## 4. Conclusion

As a non-intrusive diagnostic technique, the CCD camera can measure the spatial distributions of emission intensities. The two-dimensional profiles of electron density and electron temperature can be obtained experimentally through a CRM. Its experimental apparatus is not complex. The variations of two-dimensional distributions of plasma parameters with the RF power and the excitation frequency, as well as a mode transition from the α to the γ mode in argon discharges induced by input power were investigated experimentally. In argon plasma discharges, the axial and radial uniformities of electron density are better at a lower input power (i.e., ≤ 80 W). The axial distribution of electron density is more uniform at lower driven frequency, while the radial profile of electron temperature is smoother at a higher frequency. Generally, the electron temperature is very uniform in the bulk plasma in all the cases. Additionally the electron power absorption mode switches from the α to the γ mode with the increase of RF power at 13.56 MHz, leading to a sharp increase of electron density and an abrupt reduction of electron temperature. A spatially and temporally resolved plasma diagnostic is under consideration with the usage of ICCD. It is hoped that the spatio-temporal electron behaviors can be obtained, as a powerful diagnostic method for physics of CCPs in the future work.

## Acknowledgments

This work has been financially supported by the Fundamental Research Funds for the Central Universities (Grant No. 2232020G-10).

## ORCID iD

Xiao-jiang Huang iD https://orcid.org/0000-0003-0320-4773